# Breaking whispering-gallery modes of massless Dirac fermions in graphene quantum dots by Coulomb interaction

Zhong-Qiu Fu[1], Ke-Ke Bai[1,2], Ya-Ning Ren[1], Jiao-Jiao Zhou[3], and Lin He[1,*]

[1] Center for Advanced Quantum Studies, Department of Physics, Beijing Normal University, Beijing, 100875, People's Republic of China

[2] Institute of Physics, Hebei Normal University, Shijiazhuang, 050024, People's Republic of China

[3] Department of Mathematics and Physics, Anhui Jianzhu University, Hefei, 230601, People's Republic of China

[*]Correspondence and requests for materials should be addressed to L.H. (e-mail: helin@bnu.edu.cn).

**Coulomb interaction is of central importance in localized energy levels (bound states) or electronic flat bands and could result in many exotic quantum phases, such as magnetic, superconducting, and topological phases in graphene systems[1-14]. In graphene monolayer, the relativistic massless Dirac fermion nature of the charge carriers enables us to realize unprecedented quasibound states, which are trapped temporarily via whispering-gallery modes (WGMs) with the lifetime (trapping time) of ~ 10 fs, in circular graphene quantum dots (GQDs)[15-20]. Here we show that Coulomb interaction still plays a dominating role in determining the electronic properties of the temporarily-confined quasibound states. Our scanning tunneling microscopy (STM) and spectroscopy (STS) measurements demonstrate that the discrete quasibound state in a GQD will split into two peaks when it is partially filled. The energy separation of the two split peaks increases linear with inverse effective radius of the GQDs, indicating that the splitting arises from the Coulomb interaction. Moreover, we show that the real space distribution of the two split states separates in different regions of the GQD to reduce the Coulomb interaction, leading to the breaking of the WGM of the quasibound states.**

In localized energy levels (bound states) or electronic flat bands, the kinetic energy of the charge carriers is completely quenched or strongly suppressed. Therefore, the Coulomb interaction becomes dominant in determining electronic properties of the system, resulting in many exotic emergent quantum phases, which has been demonstrated extensively in graphene systems with the localized energy levels or the flat bands[1-14,21-26]. For example, Coulomb interaction could lift the spin degeneracy of the localized level generated by hydrogen adsorption or single-carbon defect and generate local magnetic moments around these atomic defects in graphene[1,2]. In magic-angle twisted bilayer graphene, which has low-energy flat bands, Coulomb interaction enables the realization of correlated insulator, superconductivity, ferromagnetism, and topological phases when its flat bands are partially filled[3-14]. It is well known that the kinetic energy of the charge carriers can be strongly reduced by introducing *p-n* junctions to confine them. However, owning to the massless Dirac fermions nature, the charge carriers in graphene monolayer can only be temporarily confined to form quasibound states[15-20,27-29]. In circular graphene quantum dots (GQDs), the massless Dirac fermions are temporarily confined into quasibound states via whispering-gallery modes (WGMs) because the Klein tunneling in graphene[15-20]. The quasibound states differ from both the bound states and the flat bands because they have quite short life time (trapping time) ~ 10 fs[16,27]. Then, it is natural to ask: is the Coulomb interaction still important in determining the electronic properties of the quasibound states? In this work, we show that the Coulomb interaction is very important in the quasibound states. Our experiment demonstrates explicitly that the Coulomb interaction not only splits the partial-filled quasibound states, but also breaks the WGMs of the corresponding quasibound states.

In our experiment, a new method is developed to obtain GQDs with different sizes in a continuous graphene sheet. The monolayer graphene was grown on a sulfur-rich copper foil by low pressure chemical vapor deposition (LPCVD)[30]. During the high-temperature growth process, sulfur atoms segregate from the copper foils and form self-organized sulfur islands at the interface between graphene and copper foil (see method and Fig. S1 for details of the sample preparation). According our STM experiments, the S atoms segregated from the Cu foil could form three typical structures onto the metal surface, as summarized in Fig. 1, depending on the growth temperatures of graphene. When we synthesized graphene at relative low temperature (1253 K~1273 K), the sulfur atoms tend to distribute randomly on the Cu surface as individual atoms, as shown in

Fig. 1a and 1d. With increasing the growth temperature to (1283 K~1303 K), the sulfur atoms will form well-ordered identical nanocluster superlattices, as shown in Fig. 1b and 1e. By further increasing the growth temperature to (1313 K~1333 K), the sulfur atoms will tend to assemble into nanoscale S islands, as shown in Fig. 1c and 1f. Here we should point out that the growth of graphene monolayer is not necessary for the formation of nanoscale S islands. In our experiment, we can obtain single-layer nanoscale S islands on copper substrate by annealing the sulfur-rich copper foil at high temperature (see Fig. S2). The atomic-layer difference of graphene-Cu separations between inside and outside of the nanoscale S islands introduce sharp electronic junctions and, consequently, generate nanoscale GQDs in the continuous graphene monolayer.

Figure 2a shows a representative STM topographic image of an individual GQD in a continuous graphene monolayer. The protuberance is generated by a S island in the interface between graphene and copper substrate. The inset of Fig. 2a shows height profile across the GQD and the height of the S atomic layer is measured as about 200 pm, which is consistent with that reported previously[31-33]. Atomic-resolved STM measurement (Fig. S3) demonstrates the perfect continuity of the topmost graphene lattice with no signal of defect and no intervalley scattering around the GQD. Our field-emission resonances (FER) measurements, i.e., d$z$/d$V$ spectra, indicate that the intercalation of sulfur islands introduces sharp electronic junctions between inside and outside of the nanoscale S island in the continuous graphene monolayer[28]. According to the energy shift of the first d$z$/d$V$ peak (see Fig. S4 for the details), the change of the local work function of graphene between inside and outside of the nanoscale S island is about 320 mV.

The formation of *p-n* junctions in graphene along the edges of the sulfur dots is further confirmed by carrying out dI/dV measurements. As shown in Fig. 2b, the Dirac point in the spectrum recorded outside the GQD (red curve) is about -270 meV (marked by red arrow), and the Dirac point in the spectrum measured inside the GQD (black curve) is about 50 meV (marked by black arrow). Because of the Klein tunneling at the *p-n* junction, massless Dirac fermions in graphene will be reflected from the junction with high probability at large incident angles and these reflected quasiparticles will be confined temporarily into equal-spaced quasibound states in the GQDs, which can be described by the WGM confinement[15–20]. In our experiment, a series of equally spaced quasibound states at negative energies are observed in the tunneling spectrum recorded in the GQD. Spatial-resolution STS maps, which reflect the local density of states (LDOS) in real space, show the spatial distribution of the quasibound states (Fig. 2c). The lowest quasibound state exhibits a maximum near the center of the GQD, and higher quasibound states display shell structures and progressively closer to the edge of the GQD as energy increasing, directly demonstrating the confinement via the WGMs[16-18]. Our theoretical calculations, with considering the potential height about 320 meV and the radius of GQD about 3.3 nm, as determined in the experiment, capture well the main features of the experimental results (see Fig. S5 and Supplemental Material for the calculated method), further confirming the WGM confinement of the massless Dirac fermions in the GQD. In our experiment, the WGM confinement of the massless Dirac fermions has been observed in all the studied GQDs. In graphene monolayer, the confinement of the massless Dirac Fermions is analogy to particle-in-a-box and the average energy spacing of the quasibound states can be estimated as $\Delta E = \alpha \hbar v_F / R$, where α is a dimensionless constant of order unity, $\hbar$ is the reduced Planck's constant, $v_F = 1.0 \times 10^6$ m/s is the Fermi velocity of graphene monolayer and $R$ is the effective radius of GQDs. In Fig. 2d, we summarize the energy spacing of the quasibound states $\Delta E$ as a function of the inverse effective radius ($1/R$) of the GQDs, which agrees well with the confinement of the massless Dirac fermions in the GQDs (red line in Fig. 2d).

Usually, the Coulomb interaction manifests itself mainly around the Fermi level when the bound states or the flat bands are partially filled[1-14,21-26,34-37]. Previous studies demonstrated that the Coulomb interaction will split the partial-filled bound state or flat band to lift the degeneracy of the studied system[1-14,21-26]. In our experiment, we find that the quasibound state splits into two peaks in the tunneling spectra when it is partially filled. Figure 3a shows a representative STM image of a GQD and Fig. 3b shows the corresponding STS spectra recorded in the GQD (see Fig. S6 for the spectra in a wider range of bias voltage). The energy spacing between the quasibound states, $\Delta E$ ~250 meV, agrees well with that estimated according to $\Delta E = \hbar v_F / R$ with considering the radius of the GQD $R$ ~2.5 nm, as observed in other GQDs in our experiment (Fig. 2d). However, there is an obvious difference: the first quasibound state, which is partially filled, splits into two peaks with an energy separation of about 100 meV, leaving the other fully occupied quasibound states showing no signal of splitting. We can exclude the varied electric potential as the origin of the observed phenomenon because the energy of the quasibound states is almost independent of the measured positions in this GQD, as shown by red dash line in Fig. 3b. We also can exclude coupling of the GQDs as the origin of the splitting because the studied GQD is quite isolated (see Fig. S7 for a larger STM image around the GQD) and, more importantly, the coupling of the GQDs will lead to the splitting of all the quasibound states[38]. The irregularity of the GQDs also can be excluded as the origin of the observed phenomenon because the observed splitting seems independent of the shape of the GQDs.

In our experiment, we find that the fully occupied and the completely empty quasibound states of the GQDs can be simply described by the WGM confinement and the splitting is only observed in the partial-filled quasibound states, as illustrated in Fig. 3c. The filling-related splitting in the GQDs is further confirmed in a larger GQD where the electric potential is slightly varied in the GQD, as shown in Fig. S8. One of the quasibound states changes from completely empty to fully occupied, as measured in different positions. A notable feature is that the quasibound state splits into two peaks at partial filling and does not exhibit any signal of splitting when it is empty or fully occupied. Therefore, we attribute the splitting of the partial-filled quasibound states in the GQDs to the Coulomb interaction, as observed previously in the bound states and the flat bands of graphene systems. In our experiment, the splitting of the partial-filled quasibound states is observed in different GQDs with different radii, as shown in Fig. 3d. Obviously, the splitting increases with decreasing the radius of the GQDs: the splitting increases from about 30 meV in an 8 nm GQD to about 110 meV in a 2.5 nm GQD. In Fig. 3e, we summarized the observed splitting as a function of the radius of GQDs. It is interesting to note that the splitting increases linearly with the inverse effective radius of the GQDs, which can be described quite well with considering the on-site Coulomb repulsion $U = \frac{1}{4\pi\varepsilon}\frac{e^2}{R}$. Here, $e$ is the electron charge and $\varepsilon$ is the effective dielectric constant of the GQDs. According to the linear fitting, the effective dielectric constant of the GQDs is estimated as $\varepsilon \approx (6.62 \pm 0.25)\varepsilon_0$, here $\varepsilon_0$ is vacuum dielectric constant. In previous studies[39–43], the effective dielectric constant of graphene is measured as $3\varepsilon_0 \sim 15\varepsilon_0$, depending on the supporting substrate (it is small in insulating substrate and large in conducting substrate). For simplicity, the effective dielectric constant of graphene can be estimated by $\frac{\varepsilon_0+\varepsilon_s}{2}$, where $\varepsilon_s$ is effective dielectric constant of the substrate[44]. For single layer sulfur[45], the effective dielectric constant is about $2\varepsilon_0$. In our experiment, the measured effective dielectric constant of the GQDs is larger than the estimated value because of the copper foil beneath the S islands, which enlarges the effective dielectric constant of the S islands. According to the result in Fig. 3e, the splitting generated by the Coulomb interaction will become quite small in large GQDs. For example, in a GQD with $R \sim 100$ nm, the splitting is only about 2 meV, which is almost undetectable in experiment with considering the full width of the quasibound states. Therefore, such an effect is not observed previously because the studied GQDs are quite large.

Since the Coulomb interaction splits the partial-filled quasibound states, it is expected that the Coulomb interaction also breaks the corresponding WGM of the quasibound states. Such an effect can be demonstrated explicitly by carrying out STS maps. Figure 4 shows STS maps of a GQD with the Fermi level crossing a quasibound state (Fig. S9 shows a typical STM image and representative dI/dV spectra of the GQD). The first and the second quasibound states are full-filling, whereas the third quasibound state is half filled and a splitting on the order of 30 meV is observed. For the fully occupied and completely empty quasibound states, their STS maps exhibit the expected features as that confined via the WGM in circular GQDs, as shown in Fig. 4a and 4b. However, the STS maps measured at the energies of the two split peaks of the third quasibound state show quite different features (as shown in Fig. 4c and 4d). The shell structure, as expected for the WGM confinement of the massless Dirac fermions, is clearly broken. The states for the left peak (occupied state) and the states for the right peak (empty state) distribute separately in different regions of the GQD to reduce the Coulomb interaction. Similar results have been observed in other GQDs with partial-filled quasibound states (see Fig. S10 more experimental result). Therefore, our experiment demonstrats that the Coulomb interaction not only splits the partial-filled quasibound state, but also breaks the WGM of the corresponding quasibound state in the GQD.

In conclusion, we demonstrate that the Coulomb interaction is very important in the quasibound states of the GQDs, even though the life time (trapping time) of the quasibound states is only of the order of 10 fs. To reduce the Coulomb interaction, the partial-filled quasibound state is split and, simultaneously, the WGM confinement of the quasibound states is broken. In the near future, further works should be carried out to explore possible exotic phases in the GQDs induced by the Coulomb interaction.

## Methods

**Sample preparation.** The graphene was grown by CVD method similar to previous report. The 25-μm copper foil was purchased from Alfa Aesar (46365). The copper foil was loaded into a 2-inch quartz tube of low-pressure chemical vapor deposition furnace for sample growth. The copper foil was first heated from room temperature to high temperature (1253 K-1333 K) in 300 min, with 50 sccm (standard cubic centimeter per minute) $H_2$ and 50 sccm Ar as carrier gas. Then carbon source, methane, was introduced (5 sccm) to grow graphene for 30 min. After all growth, the sample was cooled down to room temperature naturally.

**STM and STS measurements.** STM/STS characterizations were performed in ultrahigh vacuum (~$10^{-10}$ Torr) and low temperature (4 K and 77 K) scanning probe microscopes (USM-1300, USM-1400 and USM-1500) from UNISOKU. The STM tips were obtained by chemical etching from a wire of Pt/Ir (80/20%) alloys. Lateral dimensions observed in the STM images were calibrated using a standard graphene lattice and a Si (111) -(7×7) lattice and Ag (111) surface. The dI/dV measurements were taken with a standard lock-in technique by turning off the feedback circuit and using a 793-Hz 5mV A.C. modulation of the sample voltage.

## Acknowledgements

This work was supported by the National Natural Science Foundation of China (Grant Nos. 11974050, 11674029). L.H. also acknowledges support from the National Program for Support of Top-notch Young Professionals, support from "the Fundamental Research Funds for the Central Universities", and support from "Chang Jiang Scholars Program".

## Author contributions

Z.Q.F. and K.K.B. synthesized the samples and performed the STM experiments. J.J.Z. did the theoretical calculation. Z.Q.F., Y.N.R. and L.H. analyzed the data. L.H.

conceived and provided advice on the experiment, analysis, and theoretical calculation. L.H. and Z.Q.F. wrote the paper. All authors participated in the data discussion.

# Figures

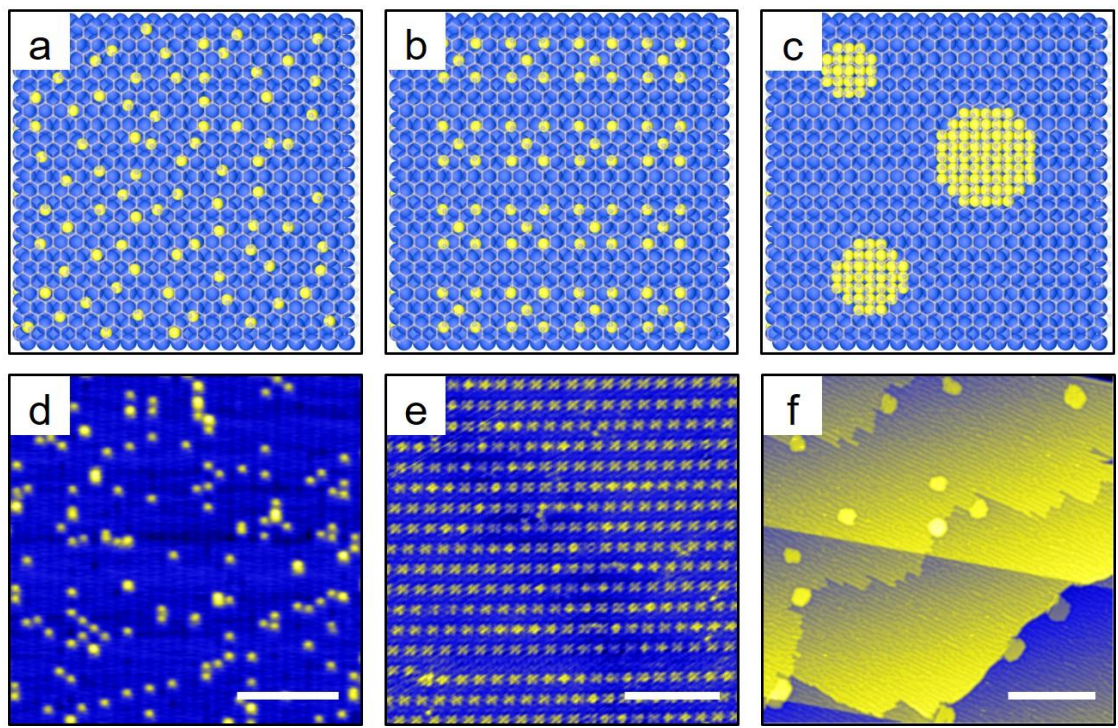

**Figure 1. Three typical structures formed by self-organized sulfur adatoms at the interface. a, d.** Schematics and STM image of individual sulfur atoms distribute randomly at the interface between graphene and copper substrate. **b, e.** Periodic sulfur nanoclusters formed at the interface. **c, f.** High density sulfur islands formed at the interface between graphene and copper substrate. Scale bars: 10 nm in **d** and **e**, 50 nm in **f**.

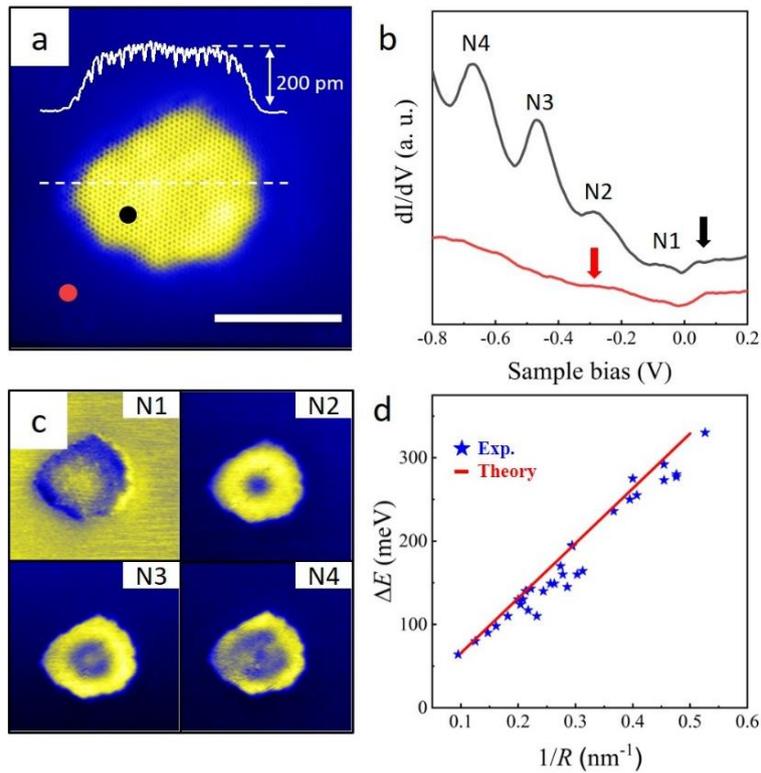

**Figure 2. Quasibound states in a GQD. a.** A STM topographic image of a GQD. Inset: Line profile along the white dash line. **b.** dI/dV spectra measured outside and inside of the GQD marked in panel **a**. **c.** STS maps recorded at the first four peaks in the dI/dV spectra of the GQD. **d.** Plot of average level spacing for several resonant peaks as a function of inverse effective radius for the GQDs. The red line is the theoretical result. Scale bar: 5 nm in **a**.

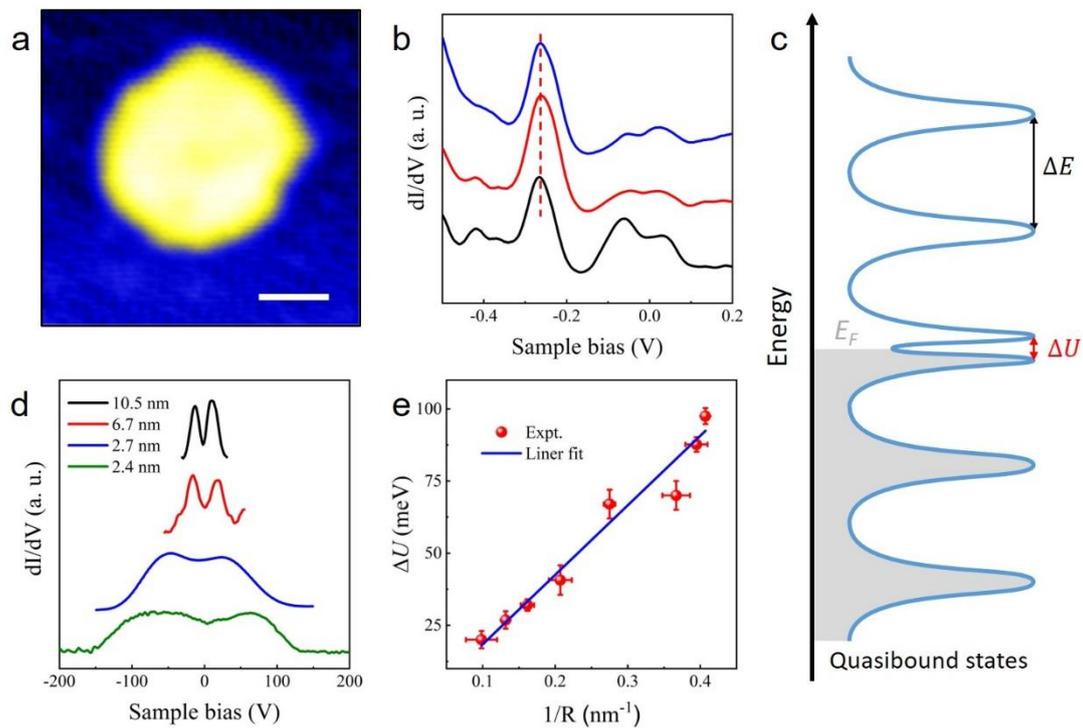

**Figure 3. Splitting of the quasibound states induced by Coulomb interaction.** A STM topographic image (**a**) and its corresponding dI/dV spectra (**b**) of a GQD. **c.** Illustration of filling-related splitting of the quasibound states. **d.** Four normalized dI/dV spectra for the GQDs with different effective radii. For clarity, the curves are offset in the y axis. **e.** Plot of average split energy as a function of inverse effective radius for the GQDs. The data is described well by a linear fit with considering Coulomb interaction in the GQDs. Scale bar: 2 nm in **a**.

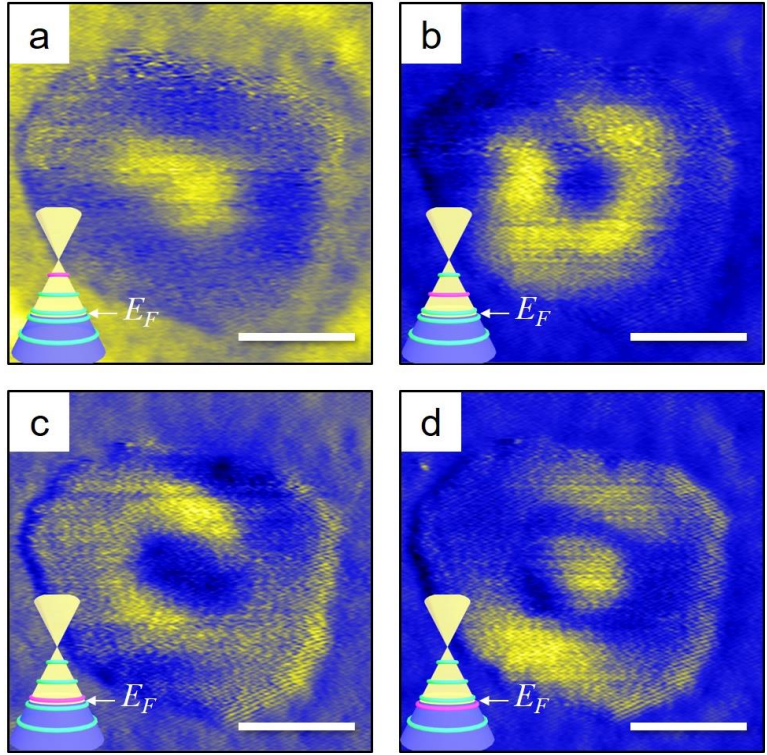

**Figure 4. Spatial distribution of LDOS recorded at the energy of empty and partial-filled quasibound states. a. b.** dI/dV mappings recorded at the energies of the empty quasibound states, as shown schematically in the inset. **c. d.** dI/dV mappings recorded at the energy of the two split peaks of the partial-filled quasibound state. Obviously, the WGM of the quasibound state is broken. Scale bars: 5 nm.